\documentclass[aps, prb, reprint, superscriptaddress, longbibliography]{revtex4-1}
\usepackage[colorlinks=true,allcolors=blue]{hyperref}
\usepackage{xcolor}
\usepackage{graphicx}
\usepackage{amsmath}
\usepackage{multirow}
\usepackage{rotating}
\usepackage{enumitem}
\usepackage{braket}
\usepackage{changes}
\usepackage{siunitx}

\begin{document} 

\title{Excited state optimization for strongly correlated quantum defects using ensemble variational Monte Carlo}

\author{Kevin G. Kleiner}
\affiliation{Anthony J. Leggett Institute for Condensed Matter Physics, Department of Physics, Grainger College of Engineering, University of Illinois at Urbana-Champaign, Urbana, Illinois 61801, USA}
\author{Lucas K. Wagner}
\affiliation{Anthony J. Leggett Institute for Condensed Matter Physics, Department of Physics, Grainger College of Engineering, University of Illinois at Urbana-Champaign, Urbana, Illinois 61801, USA}
\date{\today}

\begin{abstract} 
Using the recently introduced ensemble variational Monte Carlo (VMC), we study optimized wave functions for strongly correlated point defects, including nitrogen-vacancy and silicon-vacancy centers in diamond and substitutional iron and chromium impurities in aluminum nitride.
We study the effects of fully optimized determinant expansion parameters, orbitals, and Jastrow correlation factors on these systems. 
We find that orbitals from the hybrid functional PBE0 are much better (have lower objective functional) than semilocal PBE, which results in changes in the excitation energies up to 0.5 eV.
Further improvements can be made by directly optimizing the objective functional, resulting in changes in excitation energies up to 0.2 eV. 
The most important parameter varies from defect to defect, reinforcing the necessity of optimizing all parameters to obtain accurate excited states in strongly correlated defects.
\end{abstract}

\maketitle

\section{Introduction}

Point defects profoundly affect the properties of materials. 
In electronic devices, point defects can be beneficial, e.g., serving as dopants to tune the conductivity of semiconductors over many orders of magnitude, or detrimental, e.g., trapping charge or causing nonradiative recombination~\cite{shockleyStatisticsRecombinationsHoles1952, hallElectronHoleRecombinationGermanium1952} in optoelectronic devices.
More recently, it has been demonstrated that point defects themselves may serve as quantum devices, including qubits for quantum computing~\cite{weberQuantumComputingDefects2010, kaneSiliconbasedNuclearSpin1998}, single-photon emitters for quantum communication~\cite{aharonovichDiamondbasedSinglephotonEmitters2011, aharonovichSolidstateSinglephotonEmitters2016}, or nanoprobes for quantum metrology~\cite{schirhaglNitrogenVacancyCentersDiamond2014}. 
Recent work has expanded the study of quantum defects to transition metal~\cite{shangFirstprinciplesStudyTransition2022, leeTransitionMetalImpurities2022, otisStronglyCorrelatedStates2025, czelejTransitionMetalRelatedQuantumEmitters2024} and rare-earth~\cite{lvovskyOpticalQuantumMemory2009, kolesovOpticalDetectionSingle2012, awschalomQuantumTechnologiesOptically2018, zhangOpticalSpinCoherence2024} impurities.
In each case, dynamical operations on defects such as spin qubit initialization and readout require a detailed understanding of the defect's low-lying excited states and couplings to vibrations and neighboring nuclear spins in the host material~\cite{dohertyNitrogenvacancyColourCentre2013, goldmanStateselectiveIntersystemCrossing2015}.

\textit{Ab initio} calculations have served a key role in developing understanding of defects' electronic properties, with methods based on density functional theory (DFT) serving as the workhorse~\cite{freysoldtFirstprinciplesCalculationsPoint2014}.
While in some cases DFT has been successful in computing defect properties accurately~\cite{freysoldtFirstprinciplesCalculationsPoint2014, dreyerFirstPrinciplesCalculationsPoint2018, ivadyFirstPrinciplesCalculation2018, brobergHighthroughputCalculationsCharged2023}, strongly correlated electronic excited states of defects are a challenge, especially for transition metal defects that have significant multi-reference character~\cite{mori-sanchezManyelectronSelfinteractionError2006, maExcitedStatesNegatively2010}.
Quantum embedding methods have been developed to treat strong correlations in large, extended systems including those with defects~\cite{bockstedteInitioDescriptionHighly2018, maQuantumSimulationsMaterials2020, maFirstprinciplesStudiesStrongly2020, maQuantumEmbeddingTheory2021, muechlerQuantumEmbeddingMethods2022, shengGreensFunctionFormulation2022, otisStronglyCorrelatedStates2025}.
However, embedding methods require the user to make several difficult-to-control approximations such as the embedding subspace and treatment of screened interactions in the subspace, which can lead to qualitatively incorrect spectra for strongly correlated defects~\cite{muechlerQuantumEmbeddingMethods2022, kleinerQuantumMonteCarlo2025}. 
These challenges motivate the use of many-body methods with systematically improvable approximate excited states and the ability to scale up to extended defect systems.

In contrast to embedding methods, \textit{ab initio} QMC methods can provide a systematically improvable, fully correlated framework for defect systems based on the variational principle for ground states~\cite{foulkesQuantumMonteCarlo2001}.
QMC has been widely applied to compute ground state properties of defects, including formation energies of Schottky defects in magnesium oxide~\cite{alfeSchottkyDefectFormation2005}, self-interstitials in silicon~\cite{parkerAccuracyQuantumMonte2011}, and substitutional nitrogen defects in zinc oxide~\cite{yuFixednodeDiffusionMonte2017}.
Explicit defect excited state calculations with QMC have emerged more recently~\cite{saritasExcitationEnergiesLocalized2019, chenMulticonfigurationalNatureElectron2023, simulaCalculationEnergiesMultideterminant2023}.
However, excited state QMC results for defects can depend sensitively on the trial wave functions used: for NV$^-$:diamond, different reasonable wave function constructions have been shown to yield excitation energies varying by as much as one electron-volt~\cite{simulaCalculationEnergiesMultideterminant2023}.
Moreover, QMC has historically lacked a systematic variational criterion for assessing which trial wave functions more accurately approximate targeted excited states.
To address this gap, some of us introduced ensemble variational Monte Carlo (VMC)~\cite{wheelerEnsembleVariationalMonte2024}: a single variational principle for the lowest $N$ eigenstates of a Hamiltonian and a framework for optimizing $N$ trial states simultaneously.

In this work, we examine the use of compact multi-Slater-Jastrow wave functions built from a minimal defect active space, augmented with Jastrow correlation factors, for computing excited states of four correlated defect systems.
We perform state-specific optimization of the Jastrow, determinant expansion, and orbital parameters with the ensemble variational principle, and we assess the relative accuracies of partially and fully optimized trial states.
Under the conventional approach, in which orbital and Jastrow parameters are held fixed, trial wave functions constructed with PBE0 orbitals substantially outperform those constructed with PBE orbitals, with excitation energies changing by as much as 0.5 eV across the four defects considered.
The relative importance of further optimizing specific parameter types beyond the trial wave functions with PBE0 orbitals varies amongst the defects: some defects benefit most from state-specific orbital optimization, while others benefit most from state-specific Jastrow optimization.
Overall, full optimization beyond trial wave functions with PBE0 orbitals changes excitation energies by 0.05-0.2 eV across these systems.

\section{Defect systems} \label{section:defect_systems}

We consider the NV$^-$ and neutrally charged silicon-vacancy (SiV$^0$) centers in diamond and substitutional Fe$_{\text{Al}}^0$ and positively charged chromium (Cr$_{\text{Al}}^+$) impurities in AlN.
All four defects possess a high-spin ground state and optically addressable intra-defect excitations~\cite{daviesOpticalStudies19451976, baurDeterminationGaNAlN1994, baurPhotoluminescenceResidualTransition1995, rogersInfraredEmissionNV2008, dhaenens-johanssonOpticalPropertiesNeutral2011, greenElectronicStructureNeutral2019}, making these systems promising candidates for spin qubits and other quantum device applications.
The small supercells employed here are not meant to approximate the dilute defect concentration relevant to compare against experiment, but rather to provide a well-defined problem for assessing approximate eigenstates obtained with modest computational cost.

\begin{figure*}
\centering
\includegraphics{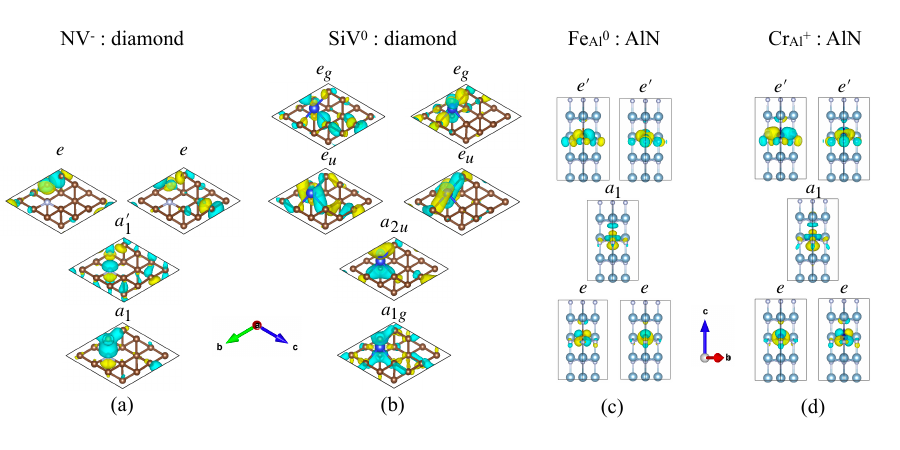}
\caption{Defect active space orbitals used for the analysis of many-body states for the (a) NV$^-$:diamond 31-atom cell, (b) SiV$^0$:diamond 31-atom cell, (c) Fe$_{\text{Al}}^0$:AlN 32-atom cell, and (d) Cr$_{\text{Al}}^+$:AlN 32-atom cell.
Each orbital is labeled by its irreducible representation of the C$_{3v}$ group for the NV$^-$:diamond, Fe$_{\text{Al}}^0$:AlN, and Cr$_{\text{Al}}^+$:AlN systems or D$_{3d}$ group for the SiV$^0$:diamond system.
For the diamond defects, brown atoms are C, grey atoms are N, and blue atoms are Si.
For the AlN defects, light blue atoms are Al, grey atoms are N, orange atoms are Fe, and blue atoms are Cr.
The orbitals were computed in Kohn-Sham DFT with the PBE0 functional.}
\label{fig:defect_orbitals} 
\end{figure*}

\subsection{NV$^-$:diamond} \label{subsection:NVdiamond}

The NV$^-$:diamond defect consists of a carbon vacancy in diamond with a substitutional nitrogen atom occupying a neighboring carbon site.
The relaxed defect geometry has C$_{3v}$ symmetry with three-fold rotation along the crystallographic [111] direction and three vertical mirror planes containing this axis~\cite{loubserElectronSpinResonance1978}.
At the spin-polarized DFT level~\cite{galiInitioSupercellCalculations2008}, the defect complex gives rise to four sp$^3$-like dangling bond orbitals from the carbon and nitrogen atoms nearest to the vacancy, occupied by six particles in the ground state.
The sp$^3$-like orbitals split into a non-degenerate a$_1$ orbital, a second non-degenerate a$_1$ orbital, and a doubly degenerate e orbital pair, labeled using irreducible representations (irreps) of the C$_{3v}$ group, which are shown in Fig.~\ref{fig:defect_orbitals}(a).
NV$^-$:diamond has a spin-triplet ground state with $^3$A$_2$ symmetry, doubly degenerate spin-singlet lowest excited states with $^1$E symmetry, a spin-singlet next lowest excited state with $^1$A$_1$ symmetry, and doubly-degenerate next lowest spin-triplet excited states with $^3$E symmetry, labeled by spin multiplicity and many-body irrep of C$_{3v}$, as determined from electron paramagnetic resonance experiments~\cite{loubserElectronSpinResonance1978}, photoluminescence experiments~\cite{daviesOpticalStudies19451976, rogersInfraredEmissionNV2008}, and various \textit{ab initio}~\cite{maExcitedStatesNegatively2010, thieringInitioCalculationSpinorbit2017, bockstedteInitioDescriptionHighly2018, bhandariMulticonfigurationalStudyNegatively2021, jinPhotoluminescenceSpectraPoint2021, barkerSpinflipBetheSalpeterEquation2022, muechlerQuantumEmbeddingMethods2022, shengGreensFunctionFormulation2022, simulaCalculationEnergiesMultideterminant2023} and phenomenological calculations~\cite{goldmanStateselectiveIntersystemCrossing2015, goldmanPhononInducedPopulationDynamics2015}.

\subsection{SiV$^0$:diamond} \label{subsection:SiVdiamond}

The SiV$^0$:diamond defect consists of a substitutional silicon atom situated equidistant in the diamond lattice between two adjacent carbon vacancy sites, forming a split vacancy configuration.
The relaxed defect geometry has D$_{3d}$ symmetry with three-fold rotation along the crystallographic [111] direction, two-fold rotation along the direction orthogonal to this axis, three diagonal mirror planes containing the three-fold rotation axis, and an inversion center at the silicon site, which maps the two vacancy sites onto each other~\cite{dhaenens-johanssonOpticalPropertiesNeutral2011}.
At the spin-polarized DFT level~\cite{galiInitioStudySplit2013}, the defect complex gives rise to six sp$^3$-like dangling bond orbitals from the carbon atoms nearest to the silicon site, occupied by ten particles in the ground state.
The sp$^3$-like orbitals split into a non-degenerate a$_{1g}$ orbital, a non-degenerate a$_{2u}$ orbital, a doubly degenerate e$_u$ orbital pair, and a doubly degenerate e$_g$ orbital pair, labeled using irreps of D$_{3d}$, which are shown in Fig.~\ref{fig:defect_orbitals}(b).
SiV$^0$:diamond has a spin-triplet ground state with $^3$A$_{2g}$ symmetry, doubly degenerate spin-singlet lowest excited states with $^1$E$_g$ symmetry, a spin-singlet next lowest excited state with $^1$A$_{1g}$ symmetry, and doubly-degenerate next lowest spin-triplet excited states with $^3$E$_u$ symmetry, as determined from electron paramagnetic resonance and photoluminescence experiments~\cite{dhaenens-johanssonOpticalPropertiesNeutral2011, greenElectronicStructureNeutral2019} and \textit{ab initio} calculations~\cite{galiInitioStudySplit2013, maQuantumSimulationsMaterials2020, maFirstprinciplesStudiesStrongly2020, maQuantumEmbeddingTheory2021}.

\subsection{Fe$_{\text{Al}}^0$:AlN} \label{subsection:FeAlN}

The substitutional Fe$_{\text{Al}}^0$:AlN defect possesses C$_{3v}$ symmetry with three-fold rotation about the crystallographic $c$-axis and three vertical mirror planes containing this axis~\cite{wickramaratneIronSourceEfficient2016, wickramaratneElectricalOpticalProperties2019, muechlerQuantumEmbeddingMethods2022}.
The four iron-nitrogen bonds form a quasi-tetrahedron with one bond roughly 0.5\% longer than the other three bonds, giving a defect geometry close to T$_d$ (tetrahedral) symmetry.
At the spin-unpolarized DFT level~\cite{wickramaratneIronSourceEfficient2016, wickramaratneElectricalOpticalProperties2019, muechlerQuantumEmbeddingMethods2022}, five iron d-like orbitals lie deep within the host band gap, occupied by five particles in the ground state. 
The d-like orbitals split into a doubly degenerate e orbital pair, a second doubly degenerate e orbital pair, and a non-degenerate a$_1$ orbital, labeled using irreps of C$_{3v}$, which are shown in Fig.~\ref{fig:defect_orbitals}(c).
Fe$_{\text{Al}}^0$:AlN has a spin-sextet ground state with $^6$A$_1$ symmetry, nearly degenerate spin-quartet lowest excited states with $^4$A$_2$ and $^4$E symmetries, and higher-lying spin-quartet excited states with $^4$A$_1$ and $^4$E symmetries, as determined from emission M\"ossbauer spectroscopy experiments~\cite{masendaLatticeSitesCharge2016}, photoluminescence experiments~\cite{baurDeterminationGaNAlN1994}, ligand field theory calculations for a d$^5$ ion in a small C$_{3v}$-symmetric crystal field compared to the interactions~\cite{suganoMultipletsTransitionMetalIons1970}, quantum embedding calculations~\cite{otisStronglyCorrelatedStates2025}, and prior QMC calculations~\cite{kleinerQuantumMonteCarlo2025}.

\subsection{Cr$_{\text{Al}}^+$:AlN} \label{subsection:CrAlN}

The substitutional Cr$_{\text{Al}}^+$:AlN defect possesses C$_{3v}$ symmetry and a one-body electronic structure qualitatively similar to Fe$_{\text{Al}}^0$:AlN.
At the spin-polarized DFT level~\cite{wuSynthesisCharacterizationModeling2003, czelejTransitionMetalRelatedQuantumEmitters2024, chinnappanFirstprinciplesStudyDefect2025}, the system has chromium d-like orbitals near the valence band maximum and in the band gap, and the d-like orbitals are occupied by two particles in the ground state.
We checked that, at the spin-unpolarized DFT level, the five d-like orbitals lie deep within the host band gap, as shown in Fig.~\ref{fig:defect_orbitals}(d), similar to the Fe$_{\text{Al}}^0$:AlN case.
Cr$_{\text{Al}}^+$:AlN has a spin-triplet ground state with $^3$A$_2$ symmetry, doubly degenerate spin-singlet lowest excited states with $^1$E symmetry, nearly degenerate higher-lying spin-triplet excited states with $^3$E and $^3$A$_1$ symmetries, and a higher-lying spin-singlet excited state with $^1$A$_1$ symmetry, as determined from photoluminescence experiments~\cite{baurPhotoluminescenceResidualTransition1995}, ligand field theory calculations for a d$^2$ ion in a large C$_{3v}$-symmetric crystal field compared to the interactions~\cite{suganoMultipletsTransitionMetalIons1970}, and prior DFT and quantum chemistry calculations~\cite{czelejTransitionMetalRelatedQuantumEmitters2024}.

\section{Method} \label{section:method}

\subsection{Geometry relaxation} \label{subsection:geometry_relaxation}

We obtain ground state equilibrium geometries for 31-atom supercells of NV$^-$:diamond and SiV$^0$:diamond and 32-atom supercells of Fe$_{\text{Al}}^0$:AlN and Cr$_{\text{Al}}^+$:AlN. 
The diamond supercells were generated from the 8-atom cubic diamond conventional cell from the Materials Project~\cite{jainCommentaryMaterialsProject2013} using an integer supercell transformation matrix with rows $(1,1,1)$, $(1,-1,1)$, and $(1,1,-1)$, which produces a 32-atom non-orthogonal supercell prior to introducing the defects.
For AlN, we generated $2\times2\times2$ supercells of the 4-atom wurtzite AlN primitive cell, also obtained from the Materials Project~\cite{jainCommentaryMaterialsProject2013}.
In each case, the respective substitutional atom or vacancy was introduced into the host supercell and the geometries were relaxed at the spin-unpolarized DFT level.
The DFT calculations used the Perdew-Burke-Ernzerhof (PBE) exchange-correlation functional~\cite{perdewGeneralizedGradientApproximation1996}, a plane wave energy cutoff of 500 eV, projector-augmented wave pseudo-potentials, a $2\times2\times2$ $k$ mesh, and Gaussian thermal smearing with width $\sigma =$ 0.01 eV.
The Fe$_{\text{Al}}^0$:AlN relaxations used \texttt{VASP}~\cite{kresseEfficiencyAbinitioTotal1996}, and the relaxations for other three defects used \texttt{Quantum Espresso}~\cite{giannozziAdvancedCapabilitiesMaterials2017}.
To check consistency between the \texttt{VASP} and \texttt{Quantum Espresso} relaxation procedures, the Fe$_{\text{Al}}^0$:AlN 32-atom cell obtained with \texttt{VASP} was relaxed using \texttt{Quantum Espresso}, and the Fe-N bond lengths were found to change by less than 0.003 \AA{}.

\subsection{Ensemble variational Monte Carlo} \label{subsection:ensemble_vmc}

To enable optimizations of multiple low-lying eigenstates in VMC, we define the ensemble objective functional
\begin{equation}
O[\{\Psi_i\}] = \sum_{i=1}^N w_i E[\Psi_i] + \lambda \sum_{i<j}^N|S_{ij}|^2
\label{eqn:ensemble_objective_functional}
\end{equation}
for an ensemble of $N$ wave functions $\{\Psi_i\} = \{\Psi_1, \Psi_2, \cdots, \Psi_N\}$, where $w_i$ is a weight on energy expectation value $E[\Psi_i]$ and $\lambda$ is a penalty on the sum of squared wave function overlaps $\sum_{i<j}^N|S_{ij}|^2$.
The ensemble variational principle~\cite{wheelerEnsembleVariationalMonte2024} states that the objective functional value for trial wave functions $\{\Psi_1, \Psi_2, \cdots\}$ is an upper bound on the objective functional value for the eigenstates $\{\Phi_1, \Phi_2, \cdots\}$, provided that all $w_i > 0$, $w_i > w_j$ for all $i, j$ where $E[\Psi_i] < E[\Psi_j]$, and
\begin{equation}
\lambda > \lambda_c = \max_{i<j} \left[ (E[\Psi_j] - E[\Psi_i]) \frac{w_iw_j}{w_i - w_j} \right].
\label{eqn:lambda_critical}
\end{equation}
The condition in Eq.~\ref{eqn:lambda_critical} penalizes the collapse of higher-energy states onto lower-energy states.

We optimize parameters in the trial wave function ensemble using gradients of the objective functional~\cite{wheelerEnsembleVariationalMonte2024}.
In the weight limit $w_j / w_i \rightarrow 0$ for all $i < j$, the objective functional gradient with respect to parameter direction $p_{j,m}$ of wave function $\Psi_j$ is approximately
\begin{align}
& \nabla_{p_{j,m}}O[\{\Psi_i\}] \nonumber \\
& \approx w_j \left(\nabla_{p_{j,m}} E[\Psi_j] + \sum_{i<j}^N 2\alpha_{ij} S_{ij}\nabla_{p_{j,m}}S_{ij}\right),
\label{eqn:objective_functional_gradient}
\end{align}
where $\alpha_{ij} > (E[\Psi_j] - E[\Psi_i]) /(1-w_j/w_i)$.
The parameter updates $\Delta p_{j,m}$ are then obtained by steepest descent in the gradient direction using the formula
\begin{equation}
\Delta p_{j,m} = -\frac{\tau}{w_j} \sum_n F_{(j,m),(j,n)}^{-1}\nabla_{p_{j,n}}O[\{\Psi_i\}],
\label{eqn:ensemble_vmc_parameter_updates}
\end{equation}
where $\tau$ is the step size rescaled by the weight $w_j$ and $F$ is the Fisher information matrix in parameter space~\cite{sorellaGeneralizedLanczosAlgorithm2001}.
These parameter updates in Eq.~\ref{eqn:ensemble_vmc_parameter_updates} are the same as those used in Ref.~\cite{entwistleElectronicExcitedStates2023}, but here they arise from the minimization of the rigorous ensemble objective in Eq.~\ref{eqn:ensemble_objective_functional}.
This strategy enables simultaneous optimization of multiple excited states within a unified ensemble VMC framework.
Monte Carlo estimates for $E[\Psi_j]$ are obtained by sampling electron coordinates from the probability distribution $\rho_j \sim |\Psi_j|^2$, and those for $S_{ij}\nabla_{p_{j,m}}S_{ij}$ are obtained by sampling from the ensemble probability distribution $\rho_{\text{ens}} \sim \sum_{i=1}^N |\Psi_i|^2$.

\subsection{Trial wave function generation} \label{subsection:trial_wfs}

We obtain VMC estimates for the \textit{ab initio} ground state and low-lying excited states at the $\Gamma$ point for each defect supercell system.
The trial wave functions have multi-Slater-Jastrow (MSJ) form,
\begin{equation}
\Psi(\vec{\alpha}, \vec{c}, \vec{\beta}) = e^{J(\vec{\alpha})}\sum_ic_i^{}D_i(\vec{\beta}),
\label{eqn:msj}
\end{equation}
where $e^{J(\vec{\alpha})}$ is a two-body Jastrow factor with parameters $\vec{\alpha}$, $\vec{c}$ are determinant expansion coefficients, and $D_i(\vec{\beta})$ is a Slater determinant constructed from one-body orbitals with parameters $\vec{\beta}$.
The determinant expansions incorporate static correlations within the defect active space orbitals, while the Jastrow factors incorporate dynamic correlations, i.e., the tendency for electrons to avoid each other at short range~\cite{umrigarOptimizedTrialWave1988, drummondJastrowCorrelationFactor2004, wagnerEnergeticsDipoleMoment2007}.
In this work, we restrict the Jastrow factor to a two-body \textit{ansatz} to minimize computational expense during wave function evaluation and optimization. 
We also restrict our calculations to \textit{minimal} determinant expansions for the defect states to keep the computational cost reasonable. 

The orbital parameters $\vec{\beta}$ in Eq.~\ref{eqn:msj} were initialized using spin-restricted open-shell DFT calculations in \texttt{PySCF}~\cite{sunSCFPythonbasedSimulations2018} using correlation-consistent pseudo-potentials~\cite{annaberdiyevNewGenerationEffective2018}, correlation-consistent valence quadruple zeta (vqz) Gaussian basis sets with decay exponents exceeding 0.1, PBE or PBE0~\cite{adamoReliableDensityFunctional1999} exchange-correlation functionals, a single $k$ point, and range-separated Gaussian density fitting~\cite{yeFastPeriodicGaussian2021}.
The DFT calculations used a target state of $S_z = +5/2$ for Fe$_{\text{Al}}^0$:AlN or $S_z = +1$ for Cr$_{\text{Al}}^+$:AlN, NV$^-$:diamond, and SiV$^0$:diamond.
The determinant expansion coefficients $\vec{c}$ were generated for each system from minimal complete active space configuration interaction (CASCI) calculations in \texttt{PySCF}, yielding the smallest determinant expansions required to represent all the low-lying excited states.
The CASCI active space for Fe$_{\text{Al}}^0$:AlN included three spin-up and two spin-down electrons in five Fe d-like DFT orbitals, that for Cr$_{\text{Al}}^+$:AlN included one spin-up and one spin-down electron in five Cr d-like DFT orbitals, that for NV$^-$:diamond included three spin-up and three spin-down electrons in four defect sp$^3$-like DFT orbitals, and that for SiV$^0$:diamond included five spin-up and five spin-down electrons in six defect sp$^3$-like DFT orbitals, which are pictured in Fig.~\ref{fig:defect_orbitals}.
Expectation values of symmetry operators evaluated with these CASCI wave functions were used to assign an irrep to each state (see Appendix~\ref{app:symmetry_classification}).
For each defect, the same fixed Jastrow factor was initialized for all states, optimized for the ground state using VMC in \texttt{PyQMC}~\cite{wheelerPyQMCAllPythonRealspace2023}.

\subsection{State-specific optimization details} \label{subsection:optimization_details}

We optimized Jastrow parameters $\vec{\alpha}$, determinant expansion coefficients $\vec{c}$, and orbital parameters $\vec{\beta}$ in MSJ trial wave functions for the lowest-lying eigenstates of each system using the ensemble VMC approach discussed in Sec.~\ref{subsection:ensemble_vmc}.
The state ensembles included the lowest-lying six states for NV$^-$:diamond and SiV$^0$:diamond and the lowest-lying seven states for Fe$_{\text{Al}}^0$:AlN and Cr$_{\text{Al}}^+$:AlN.

To assess the relative importance of optimizing each subset of wave function parameters, we compare decreases to the ensemble objective functional from a sequence of optimization stages.
We begin the optimization with $\vec{\beta}$ computed from DFT using the widely used PBE exchange-correlation functional, $\vec{\alpha}$ optimized for the ground state and fixed for all states, and $\vec{c}$ computed from CASCI using active spaces of PBE orbitals.
We then generated the wave functions with the same procedure, except using orbitals generated with the PBE0 functional in DFT~\cite{adamoReliableDensityFunctional1999}, which assesses the variational improvement gained by generating $\vec{\beta}$ (and hence $\vec{c}$ and $\vec{\alpha}$) with a hybrid DFT functional rather than a semi-local DFT functional.
Starting from the PBE0 orbital parameters, CASCI determinant coefficients, and fixed Jastrow parameters, we optimized $\vec{\alpha}$, then optimized $\vec{\alpha}$ and $\vec{c}$ together, then optimized $\vec{\alpha}$, $\vec{c}$, and $\vec{\beta}$ together.
The Jastrow optimizations included on the order of 100 $\vec{\alpha}$ parameters per state, the determinant expansion optimizations added on the order of 10 $\vec{c}$ parameters per state, and the orbital optimizations added the 200 largest $\vec{\beta}$ parameters in absolute value per defect orbital per state, which reduces computational expense and memory requirements.
This sequence progressively enlarges the parameter space used in optimization, which leads to equal or lower objective functional minima under sufficiently converged optimizations.
The parameter updates defined in Eq.~\ref{eqn:ensemble_vmc_parameter_updates} used $\tau = 0.05$, which proved sufficiently small to avoid divergence of the objective functional during optimization.

We compute the ensemble objective functional value in Eq.~\ref{eqn:ensemble_objective_functional} for the trial states in each optimization stage.
The weights $w_i$ on the energy expectation values were linearly spaced between 1 and 0.05, then normalized, which satisfies the ordering and positivity conditions required by the ensemble variational principle.
Because the optimization was performed in the limit $w_j / w_i \rightarrow 0$ for all $i < j$, as discussed in Sec.~\ref{subsection:ensemble_vmc}, weight values were used only for evaluating the ensemble objective functional.
Different admissible weight choices would alter the numerical value of the objective functional, but are not expected to qualitatively change the trends reported here.
The overlap penalty $\lambda$ was set to 10 eV, which exceeds the critical value $\lambda_c \leq 5.6$ eV evaluated from Eq.~\ref{eqn:lambda_critical} for all cases considered.
To obtain Monte Carlo averages for the energies and overlaps, we discarded the first block as equilibration and treated the remaining blocks as quasi-independent.

Because the accuracy of VMC-computed energies is limited by the flexibility of the variational \textit{ansatz}, we additionally perform fixed-node diffusion Monte Carlo (FN-DMC) calculations using the optimized trial states to assess the impact of the approximate \textit{ansatz} on the excitation energies.
FN-DMC projects the trial wave function in imaginary time while constraining its nodal surface to remain fixed, yielding the lowest-energy wave function with the same nodes as the trial state~\cite{foulkesQuantumMonteCarlo2001}.
FN-DMC provides a variational upper bound on the exact eigenstate energy for states belonging to one-dimensional irreps such as A$_1$ and A$_2$, but the bound does not generally hold for states belonging to multi-dimensional irreps such as E~\cite{foulkesSymmetryConstraintsVariational}, although FN-DMC becomes exact if the node surface matches that of the target eigenstate.
We performed FN-DMC calculations with each fully optimized trial state at the time steps $\delta \tau = \{0.005, 0.01, 0.015, 0.02, 0.025\}$ Ha$^{-1}$.
The total energies were extrapolated to zero time step using weighted least squares regression.
In the Monte Carlo averaging for the energies, we discarded the first 30 samples as equilibration and reblocked the remaining samples into 10 quasi-independent blocks.

\section{Results and discussion} \label{section:results_discussion}

We first examine decreases in the ensemble objective functional across successive stages of wave function optimization.
We then analyze how these optimization stages affect the computed excitation energies and finally assess the changes in excitation energies associated with FN-DMC projection on the fully optimized trial states.
The objective functional optimizations, VMC-computed excitation energy spectra during optimization, excitation energy changes from optimization, and excitation energy changes from FN-DMC projection are presented in Figs.~\ref{fig:objective_functional_optimization}, \ref{fig:excitation_energy_levels_during_optimization}, \ref{fig:excitation_energy_changes}, and \ref{fig:dmc_excitation_energy_levels}, respectively.

\subsection{State-specific optimization} \label{subsection:state_specific_optimization}

\begin{figure}
\centering
\includegraphics{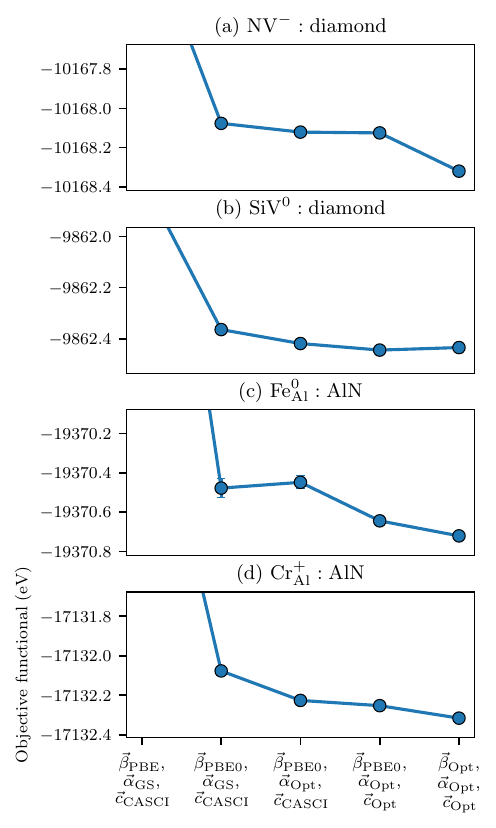}
\caption{Optimization of the ensemble objective functional with respect to Jastrow parameters $\vec{\alpha}$, determinant expansion coefficients $\vec{c}$, and orbital parameters $\vec{\beta}$ in MSJ trial wave functions for (a) NV$^-$:diamond, (b) SiV$^0$:diamond, (c) Fe$_{\text{Al}}^0$:AlN, and (d) Cr$_{\text{Al}}^+$:AlN.
The objective functional means and statistical error bars are shown in eV units.
The panels denote the parameters $\vec{\beta}$, $\vec{\alpha}$, and $\vec{c}$ in each stage of optimization.
The y-axis scale only includes objective functional values for the $\vec{\beta}_{\text{PBE0}}$, $\vec{\alpha}_{\text{GS}}$, and $\vec{c}_{\text{CASCI}}$ case and further optimization.}
\label{fig:objective_functional_optimization}
\end{figure}

In Fig.~\ref{fig:objective_functional_optimization}, we show the systematic improvement of the objective functional.
For all defects and optimization stages, the sum of squared overlaps was $\lesssim$~0.001, so that the overlap penalty term in the objective functional contributes $\lesssim$~0.01 eV.
The dominant variational improvements come from using the PBE0 functional rather than PBE to generate the orbital parameters $\vec{\beta}$, mirroring early results on the ground state of transition metal systems~\cite{wagnerQuantumMonteCarlo2003, kolorencWaveFunctionsQuantum2010}.
By optimizing the variational parameters using state-specific optimization, we can remove dependence on the initial parameter guesses from DFT and/or configuration interaction.

Overall, we find that variationally optimizing the wave function parameters results in a decrease in the ensemble objective functional of up to 0.2 eV compared to the PBE0-based \textit{ansatz}.
Which parameters are most important varies amongst the defects. 
The $\vec{\alpha}$ optimization leads to the largest observed decrease in objective functional for Cr$_{\text{Al}}^+$:AlN, the $\vec{\alpha}$ and $\vec{c}$ optimization leads to the largest decrease for Fe$_{\text{Al}}^0$:AlN, and the $\vec{\beta}$, $\vec{\alpha}$, and $\vec{c}$ optimization leads to the largest decrease for NV$^-$:diamond, as shown in Fig.~\ref{fig:objective_functional_optimization}.
SiV$^0$:diamond shows comparatively smaller changes in objective functional across all optimization stages, and the functional stays the same within error bars during the $\vec{\alpha}$, $\vec{c}$, and $\vec{\beta}$ optimization. 
Overall, the variation across defects in which parameter classes contribute most strongly to variational improvements highlights the importance of state-specific optimization of all parameter classes.

\subsection{Effects of optimizing each parameter set on the spectrum} \label{subsection:validation_of_spectra}

\begin{figure}
\centering
\includegraphics{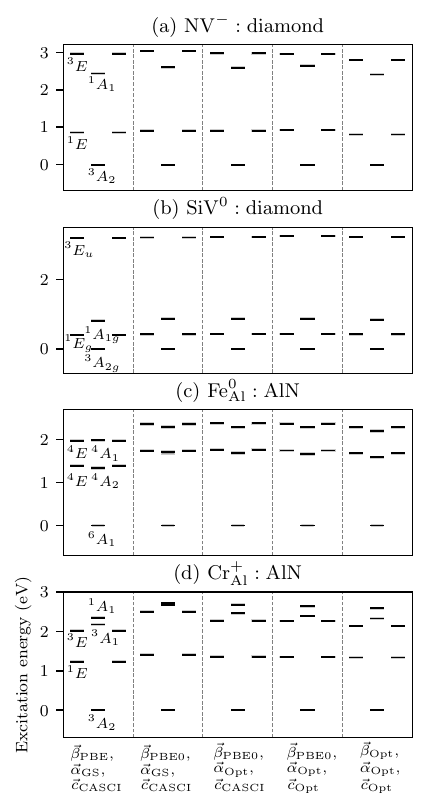}
\caption{VMC-computed excitation energies relative to the ground state energy at each respective optimization stage.
The energy levels are labeled according to the state's spin degeneracy and many-body symmetry irrep.
Single-$\sigma$ statistical errors on the excitation energies are shown as light-shaded grey boxes, mostly thinner than the level thickness.
Doubly-degenerate $E$ states are horizontally displaced relative to the non-degenerate states.}
\label{fig:excitation_energy_levels_during_optimization}
\end{figure}

In Fig.~\ref{fig:excitation_energy_levels_during_optimization}, we show the excitation energies of the lowest-lying $N$ states for each defect, as a function of the parameters optimized. 
We note that ensemble optimization is not guarunteed to systematically improve the excitation energies except when the trial states are very close to the exact eigenstates, which is not guarunteed with the minimal MSJ parameterizations used in this work.
Given this caveat, it is striking that the energy ordering of the trial states by symmetry irrep is robust to optimization, even though the decreases in the total energies can be quite large, in particular between using PBE and PBE0 orbitals in the \textit{ansatz}. 

As one might expect from other studies on transition metal systems~\cite{kolorencWaveFunctionsQuantum2010}, the transition metal defects are more sensitive to the orbitals used, with the PBE0 orbitals changing the excitation energies of the Fe$_{\text{Al}}^0$:AlN and Cr$_{\text{Al}}^+$:AlN defects by multiple tenths of an eV. 
Optimizing the orbitals beyond that results in much smaller changes, similar to previous findings by one of us in CaCuO$_2$~\cite{chowCapturingSpinFluctuations2024}.

\subsection{Effects of optimization on excitation energies} \label{subsection:optimization_on_excitation_energies}

\begin{figure}
\centering
\includegraphics{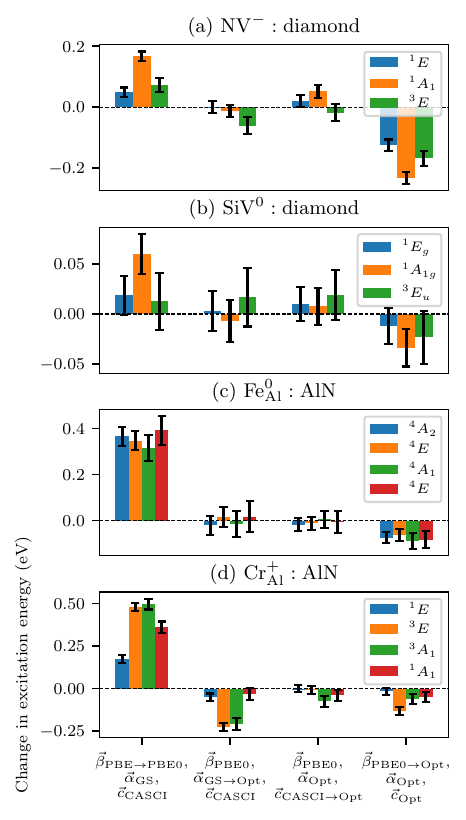}
\caption{Changes in VMC-computed excitation energies during optimization.
The bars denote the excited state labeled by symmetry irrep.
One state per doubly-degenerate $E$ irrep is shown.
Single-$\sigma$ statistical errors are shown at the peak of each bar.
Zero change in excitation energy is denoted with a horizontal dashed line.}
\label{fig:excitation_energy_changes}
\end{figure}

In Fig.~\ref{fig:excitation_energy_changes}, we show a zoomed-in plot of the change in excitation energies for each state. 
By comparing to Fig.~\ref{fig:objective_functional_optimization}, we can determine the effects of error cancellation on the spectrum. 
For the PBE $\rightarrow$ PBE0 improvement in the orbitals, there is a large amount of error cancellation, since the total energies decrease by multiple eV, while the excitation energies only change by 0.05-0.5 eV.
Beyond the PBE0-based \textit{ansatz}, we find no discernible trends in error cancellation. 
In the case of orbital optimization for NV$^-$:diamond, there is little error cancellation, indicating that the optimal orbitals are different for the excited states than for the ground state. 
This interpretation is consistent with the state-dependent changes in the one-body reduced density matrices within the defect active space for NV$^-$:diamond (see Appendix~\ref{app:optimization_on_rdms}).
On the other hand, optimizing the $\vec{c}$ coefficients for Fe$_{\text{Al}}^0$:AlN resulted in a large change in the objective functional but little change in excitation energies. 
Interestingly, for Cr$_{\text{Al}}^+$:AlN, the optimal Jastrow correlation factor is different for different states, in particular the triplet versus the singlet states.

Fig.~\ref{fig:excitation_energy_changes} also shows variation across defects in which optimization stages most strongly affect the excitation energies.
The use of PBE0 orbitals rather than PBE orbitals generally increases the excitation energies, with larger increases for Fe$_{\text{Al}}^0$:AlN and Cr$_{\text{Al}}^+$:AlN than for NV$^-$:diamond and SiV$^0$:diamond.
Further variational improvements beyond the PBE0-based trial states generally lead to negative shifts in excitation energies: $\vec{\alpha}$ optimization leads to the largest decreases for Cr$_{\text{Al}}^+$:AlN, while the combined $\vec{\alpha}$, $\vec{c}$, and $\vec{\beta}$ optimization leads to the largest decreases for NV$^-$:diamond and Fe$_{\text{Al}}^0$:AlN.

\subsection{Accuracy checks}

\begin{figure}
\centering
\includegraphics{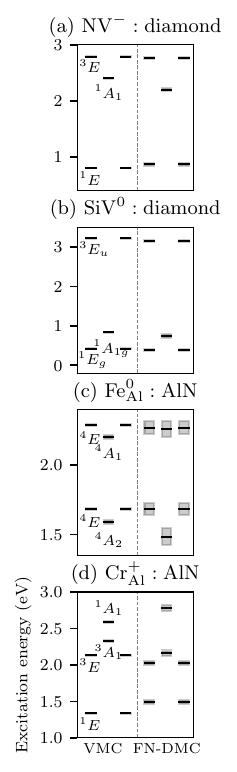}
\caption{Comparison of defect excitation energies computed with VMC and FN-DMC using the fully optimized trial states.
The ground states at zero excitation energy are omitted for clarity.}
\label{fig:dmc_excitation_energy_levels}
\end{figure}

Before comparing to other results, let us first go over the caveats in such a comparison. 
This work is meant mainly to study the effects of optimizing variational parameters in variational Monte Carlo on estimated excitation energies of several defects, and as a test that such calculations are possible. 
To do this, we necessarily have performed the calculations at small unit cells and neglecting some approximations. 
These approximations include spin-orbit coupling, 
the small system sizes considered here, effective core potentials, vertical versus zero-phonon-line (ZPL) energy, and the variational form of the wave function, including basis sets, Jastrow, and determinant expansion. 

The spin-orbit coupling in the transition metals is very small~\cite{czelejTransitionMetalRelatedQuantumEmitters2024} and much smaller for the diamond defects. 
The difference between the vertical excitation energy and the ZPL energy varies by defect. 
This can be estimated by the difference between the absorption maximum and the ZPL energy, as collected in Table~\ref{tab:experimental_summary}.
For NV$^-$:diamond, this energy difference is around 0.2 eV, while for the SiV$^0$:diamond and Cr$_{\text{Al}}^+$:AlN defects, the relaxation is small. 

We have estimated the finite size approximation by studying 72-atom supercells of Fe$_{\text{Al}}^0$:AlN and Cr$_{\text{Al}}^+$:AlN, the full details of which are available in Ref.~\cite{kleinerQuantumMonteCarlo2025a}.
We find a difference of approximately 0.4 eV between these cells, mainly increasing the excitation energy, so they are clearly a dominant source of difference from experiment, and we may not be in the asymptotic limit for these small cells.

Finally, we assess how much of the error in the excitation energies can be attributed to the approximate variational \textit{ansatz} by performing FN-DMC projection on the fully optimized trial states, as shown in Fig.~\ref{fig:dmc_excitation_energy_levels}.
Across the four systems, FN-DMC changes the excitation energies by up to a couple tenths of an eV relative to those obtained with VMC.
In contrast, FN-DMC decreases the total energies by on the order of 10 eV under imaginary time projection.
The relatively small changes in excitation energies show that much of the errors from the variational \textit{ansatz} cancel in energy differences. 
The remaining errors in the energy differences are comparable in magnitude to the other sources of error considered here.

In conclusion, from the above, we should not be surprised by differences of 0.5 eV or larger from reference infinite-cell, ZPL values. 
As shown in the appendix tables, we find differences of similar magnitude to this with several notable exceptions: the $^3$E$_u$ state in the SiV$^0$:diamond defect and the $^1$A$_1$ and $^3$E states in the NV$^-$:diamond defect, which are overestimated by $\sim$1-2 eV. 
Simula estimated the finite size effects on these states, and found it to be small compared to that difference~\cite{simulaCalculationEnergiesMultideterminant2023}.
These states might have substantial weight on the host bands with delocalized character, which would mean that more determinants should be included in the \textit{ansatz} considered here to obtain high accuracy. 
This would be an interesting future study.

It is worth noting that the accuracy is quite similar between the transition metal defects and the diamond defects. 
It appears that the current technique is more robust than embedding techniques, which can fail catastrophically for defects like Fe$_\textrm{Al}^0$:AlN~\cite{muechlerQuantumEmbeddingMethods2022}.

\section{Conclusion} \label{section:conclusion}

In summary, we applied the recently introduced ensemble variational Monte Carlo (VMC) methodology~\cite{wheelerEnsembleVariationalMonte2024} to optimize compact multi-Slater-Jastrow wave functions for excited states of four correlated defects in small supercells, including nitrogen-vacancy and silicon-vacancy centers in diamond and iron and chromium impurities in aluminum nitride.
Ensemble VMC provides a rigorous variational framework for ensembles of states, enabling optimization for the lowest $N$ target eigenstates of a Hamiltonian.
Using ensemble VMC, we were able to simultaneously optimize thousands of state-specific variational parameters in multi-Slater-Jastrow (MSJ) wave functions for the defects, including their two-body Jastrow factors, determinant expansion coefficients, and orbitals.
With this approach, we achieved excitation energies that reproduce the experimental ordering, with higher excitation energies overestimated by an amount consistent with the small supercells, entirely from first principles.

The cost of optimizing the orbitals is rather high, which limited the size of system we could study. 
We found that the improvement obtained by optimizing the orbitals, at least compared to using almost-optimal orbitals from PBE0, is small compared to the errors induced by using a very small unit cell and small determinant expansion to make the calculation attainable. 
On the other hand, optimizing the Jastrow and determinant coefficients is quite inexpensive and offers improvements for some defects. 
If one wishes to study the defects in the dilute limit, it may well be more effective to vary the DFT functional and use the ensemble variational principle to select the best orbitals, which can be done very inexpensively. 
Such an approach would allow one to access much larger supercells than considered here.

\section{Author Contributions}

\textbf{Kevin G. Kleiner} performed the quantum Monte Carlo calculations and wrote the original draft of the manuscript.
\textbf{Lucas K. Wagner} conceptualized the project, aided in analyzing the excited state results, contributed to the preparation of figures, contributed to writing and revising the manuscript, and supervised the project.

\begin{acknowledgments}
We thank William Wheeler for helpful discussions about generating small non-orthogonal supercells of diamond.
KGK and LKW acknowledge support from the U.S. Department of Energy (DOE), Office of Science, Office of Basic Energy Sciences, Computational Materials Sciences Program, under Award No. DE-SC0020177, which supported KGK in the quantum Monte Carlo calculations and the writing of the original manuscript and LKW in supervising and writing and revising the manuscript.
KGK and LKW acknowledge computing resources from the Illinois Campus Cluster Program, which supported quantum Monte Carlo calculations.
KGK acknowledges support from the NSF Graduate Research Fellowship Program under Award No. DGE-1922758, which supported the initial quantum Monte Carlo calculations and the writing of the original manuscript.
An award for computer time was provided by the U.S. DOE's Innovative and Novel Computational Impact on Theory and Experiment (INCITE) Program, which supported quantum Monte Carlo calculations.
This research used resources from the Argonne Leadership Computing Facility, a U.S. DOE Office of Science user facility at Argonne National Laboratory, which is supported by the Office of Science of the U.S. DOE under Contract No. DE-AC02-06CH11357.
\end{acknowledgments}

\section{Data availability}

Data\cite{kevinkleiner2026data} is provided at the Materials Data Facility\cite{blaiszikDataEcosystemSupport2019,blaiszikMaterialsDataFacility2016}.

\appendix

\section{Wave function symmetry classifications} \label{app:symmetry_classification}

In cases where the electronic Hamiltonian commutes with symmetry operators in a point group, the Hamiltonian and the symmetry operators share a common set of eigenstates.
When the Hamiltonian is diagonalized using an approximate method such as CASCI (and the active space orbitals used are symmetry adapted), the resulting approximate eigenstates can typically be labeled using irreps of the point group.

We obtain irrep labels for the CASCI approximate eigenstates using expectation values of symmetry operator $\hat{R}$ defined as
\begin{align}
\langle \Psi|\hat{R}|\Psi \rangle & = \sum_{kk'}c_k^*c_{k'}\langle D_k^{\uparrow}D_k^{\downarrow}|\hat{R}|D_{k'}^{\uparrow}D_{k'}^{\downarrow} \rangle \nonumber \\
& = \sum_{kk'}c_k^*c_{k'}\text{det}[\langle \phi_{a_k}^{\uparrow}|\hat{R}|\phi_{b_{k'}}^{\uparrow}\rangle]\text{det}[\langle \phi_{a_k}^{\downarrow}|\hat{R}|\phi_{b_{k'}}^{\downarrow}\rangle],
\label{symmetry_expectation_value}
\end{align}
where $\{\phi_{a_k}^{\sigma}\}$ and $\{\phi_{b_{k'}}^{\sigma}\}$ denote the sets of spin-$\sigma$ orbitals in determinants $D_k^{\sigma}$ and $D_{k'}^{\sigma}$ respectively.
In obtaining the final line of Eq.~\ref{symmetry_expectation_value}, we used the property that the overlap of two Slater determinants is given by the determinant of overlaps between their respective orbitals~\cite{plasserEfficientFlexibleComputation2016}.

The irrep labels are then assigned by comparing the computed expectation values (characters) to those in the character table of the corresponding point group.
The NV$^-$:diamond, Fe$_{\text{Al}}^0$:AlN, and Cr$_{\text{Al}}^+$:AlN defects studied in this work belong to the C$_{3v}$ group, which contains three-fold rotations $\hat{C}_3$ and vertical mirror reflections $\hat{\sigma}_v$, with irreps A$_1$, A$_2$, and E. 
The SiV$^0$:diamond defect belongs to the D$_{3d}$ group, which includes $\hat{C}_3$, two-fold rotations $\hat{C}_2'$ about axes perpendicular to those of $\hat{C}_3$, diagonal mirror reflections $\hat{\sigma}_d$, six-fold improper rotations $\hat{S}_6$, and inversion $i$, with irreps A$_{1g}$, A$_{1u}$, A$_{2g}$, A$_{2u}$, E$_g$, and E$_u$.
This assignment procedure follows that used in Ref.~\cite{muechlerQuantumEmbeddingMethods2022}.
To check whether the Jastrow factor significantly alters the symmetry character, we computed $\langle \Psi|\hat{C}_3|\Psi\rangle$ and $\langle \Psi|\hat{\sigma}_v|\Psi \rangle$ for the lowest few multi-Slater-Jastrow approximate eigenstates with fixed two-body Jastrow factors for Fe$_{\text{Al}}^0$:AlN in variational Monte Carlo. 
The resulting expectation values differ only modestly from the corresponding CASCI values, indicating that the symmetry assignments for CASCI are robust.

We determine spin degeneracies for the CASCI eigenstates using the expectation value of $\hat{S}^2$ written in second quantization as 
\begin{align}
\langle \Psi|\hat{S}^2|\Psi \rangle \nonumber
& = \sum_{i,j}^{N_{\text{orb}}} \langle \Psi|\hat{\mathbf{S}}_i \cdot \hat{\mathbf{S}}_j|\Psi \rangle \nonumber \\
& = \frac{1}{2}(N_{\text{orb}} + 2)\sum_i^{N_{\text{orb}}}\sum_{\sigma \in \{\uparrow,\downarrow\}}\langle \Psi|\hat{n}_{i,\sigma}|\Psi \rangle \nonumber \\
& - \frac{1}{2}\sum_{i,j}^{N_{\text{orb}}}\sum_{\sigma,\sigma' \in \{\uparrow,\downarrow\}}\langle \Psi|\hat{c}_{i,\sigma}^{\dagger}\hat{c}_{j,\sigma}^{}\hat{c}_{j,\sigma'}^{\dagger}\hat{c}_{i,\sigma'}^{}|\Psi\rangle \nonumber \\
& - \frac{1}{4}\sum_{i,j}^{N_{\text{orb}}}\sum_{\sigma,\sigma' \in \{\uparrow,\downarrow\}}\langle \Psi|\hat{n}_{i,\sigma}\hat{n}_{j,\sigma'}|\Psi \rangle,
\label{spin_squared}
\end{align}
where the sum includes the $N_{\text{orb}}$ orbitals in the active space. 
The expression for spin squared in Eq.~\ref{spin_squared} is consistent with that derived in Ref.~\cite{couryHubbardlikeHamiltoniansInteracting2016}.

\section{Effects of the Jastrow factor on excitation energies} \label{app:jastrow_on_excitation_energies}

\begin{figure}
\centering
\includegraphics{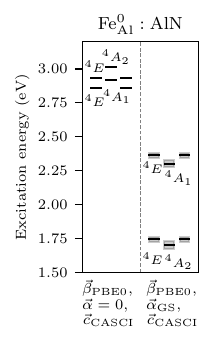}
\caption{Comparison of QMC-computed excitation energies for Fe$_{\text{Al}}^0$:AlN between MSJ trial states with and without Jastrow parameters $\vec{\alpha}_{\text{GS}}$.
Both sets of wave functions have $\vec{\beta}_{\text{PBE0}}$ orbital parameters and $\vec{c}_{\text{CASCI}}$ determinant coefficients.
The $^6$A$_1$ ground state is omitted for clarity.}
\label{fig:casci_vs_casci_jastrow_levels}
\end{figure}

In Fig.~\ref{fig:casci_vs_casci_jastrow_levels}, we show the differences in computed excitation energies for Fe$_{\text{Al}}^0$:AlN between MSJ trial states with and without Jastrow parameters $\vec{\alpha}_{\text{GS}}$.
The $\vec{\alpha} = 0$ wave functions correspond to CASCI eigenstates constructed from the PBE0 orbital parameters.
Appending the ground state Jastrow factor to the trial states lowers the excitation energies by roughly 1 eV and reverses the ordering of the lowest $^4$A$_1$ and $^4$A$_2$ states relative to CASCI alone.
These excitation energy shifts and changes in state ordering demonstrate the critical role of the Jastrow correlation factor in obtaining qualitatively correct spectra for Fe$_{\text{Al}}^0$:AlN when using compact MSJ wave functions.

\section{Effects of optimization on reduced density matrices} \label{app:optimization_on_rdms}

\begin{figure}
\centering
\includegraphics{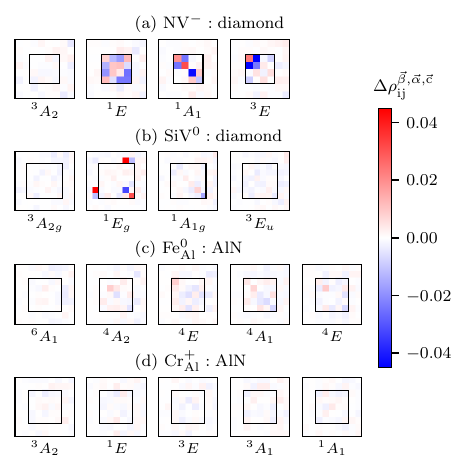}
\caption{Changes in computed one-body reduced density matrices for each trial state from the $\vec{\beta}$, $\vec{\alpha}$, and $\vec{c}$ optimization relative to those from $\vec{\alpha}$ and $\vec{c}$ optimization.
Each box represents one many-body state, labeled by irrep.
One state per doubly-degenerate $E$ irrep is shown.
The inner boxes include changes in the defect active space orbitals pictured in Fig.~\ref{fig:defect_orbitals}, and the outer boxes include changes in the Kohn-Sham bulk orbitals near the band edges. 
Single-$\sigma$ statistical errors in the density matrix changes are on the order of 0.006 or smaller.}
\label{fig:rdm1_changes_orbital_optimization}
\end{figure}

In Fig.~\ref{fig:rdm1_changes_orbital_optimization}, we show the changes in one-body reduced density matrices $\rho_{ij} = \langle\Psi|\hat{c}_i^{\dagger}\hat{c}_j|\Psi\rangle$ for each trial state $\Psi$ after $\vec{\beta}$, $\vec{\alpha}$, and $\vec{c}$ optimization relative to those with optimized $\vec{\alpha}$ and $\vec{c}$ parameters.
The density matrix changes characterize the extent of orbital relaxations in each state relative to the Kohn-Sham orbitals.
We observe significant relaxation of defect orbitals in the excited states of NV$^-$:diamond, whereas SiV$^0$:diamond, Fe$_{\text{Al}}^0$:AlN, and Cr$_{\text{Al}}^+$:AlN generally exhibit smaller changes.
The apparent orbital relaxations in the $^1$E$_g$ excited states of SiV$^0$:diamond primarily reflect arbitrary rotations within degenerate subspaces: the occupancy changes in the two e$_g$ orbitals nearly cancel, and the off-diagonal defect-bulk matrix element changes nearly cancel between the two $^1$E$_g$ states, only one of which is shown.
For Fe$_{\text{Al}}^0$:AlN and Cr$_{\text{Al}}^+$:AlN, the density matrix changes are barely larger than statistical noise and are distributed across both defect and bulk orbitals, which we believe reflects the changes in the $\vec{\alpha}$ and $\vec{c}$ parameters during optimization.
These density matrix changes are consistent with the excitation energy changes from orbital optimization shown in Fig.~\ref{fig:excitation_energy_changes}.

\section{Literature and experimental evaluations}

% Experimental summary table: raw measured quantities (absorption max, ZPL,
% emission max) for the four defects studied in this work. Vertical
% absorption max is the appropriate comparison for the computed vertical
% excitation energies in this work. Entries with "$\sim$" or "---" reflect
% the level of available experimental detail (see Notes).

\begin{table*}
\caption{Experimental energies (eV) for the lowest spin-allowed optical
  transition of each defect. The vertical absorption maximum is the
  appropriate experimental comparison for vertical excitation energies
  computed in this work; for defects with a high Debye--Waller factor the
  absorption and emission maxima are close to the ZPL.}
\label{tab:experimental_summary}
\begin{ruledtabular}
\begin{tabular}{lcccc}
Defect & Transition & Absorption max & ZPL & Emission max \\
\hline
NV$^-$:diamond~\cite{daviesOpticalStudies19451976,rogersInfraredEmissionNV2008} & $^3A_2 \to {}^3E$ & $\sim 2.18$ & 1.945 & $\sim 1.76$ \\
SiV$^0$:diamond~\cite{dhaenens-johanssonOpticalPropertiesNeutral2011,greenElectronicStructureNeutral2019} & $^3A_{2g} \to {}^3E_u$ & $\approx$~ZPL & 1.31 & $\approx$~ZPL \\
Fe$_{\mathrm{Al}}^0$:AlN~\cite{baurDeterminationGaNAlN1994,masendaLatticeSitesCharge2016} & $^6A_1 \to {}^4T_1$ & --- & 1.299 & --- \\
Cr$_{\mathrm{Al}}^+$:AlN~\cite{baurPhotoluminescenceResidualTransition1995} & $^3A_2 \to {}^3E$ & $\approx$~ZPL & 1.193 & $\approx$~ZPL \\
\end{tabular}
\end{ruledtabular}
\textit{Notes.} For NV$^-$:diamond, the absorption and emission band maxima are
read from the broad sidebands in Davies \& Hamer~\cite{daviesOpticalStudies19451976};
total Stokes shift is $\sim 0.42$~eV.
SiV$^0$:diamond has a Debye--Waller factor of $\sim 0.9$~\cite{greenElectronicStructureNeutral2019},
so $\geq 90\%$ of the emission is in the ZPL itself; a phonon replica
appears at $\sim 1.27$~eV (39~meV below the ZPL, dominated by Si
vibration), but the band maxima are essentially at the ZPL.
For Fe$_{\mathrm{Al}}^0$:AlN, only the ZPL is sharply identified in
Ref.~\cite{baurDeterminationGaNAlN1994}; the emission is a broad
sideband extending to lower energies whose maximum is not separately
tabulated in the primary reference.
Otis~et~al.~\cite{otisStronglyCorrelatedStates2025} compute an
excited-state relaxation energy of 0.08~eV (computed with time-dependent DFT with the HSE functional), implying a
vertical absorption near 1.38~eV.
For Cr$_{\mathrm{Al}}^+$:AlN, the ZPL is sharp with a modest phonon replica
$\sim 75$~meV below~\cite{czelejTransitionMetalRelatedQuantumEmitters2024},
so both band maxima are close to the ZPL.
The $^4T_1$ irrep splits into $^4A_2 + {}^4E$ under $C_{3v}$ for
Fe$_{\mathrm{Al}}^0$:AlN.
\end{table*}

% Compilation tables of experimental and prior-theory excitation energies
% for the four defects studied in this work, with this work's VMC values
% from 3_data/qmc_descriptors_all.csv (geminal=False rows).
% All energies in eV.

\begin{table*}
\caption{Vertical excitation energies (eV) for NV$^-$:diamond, relative to the
  $^3A_2$ ground state. This work's values are from a 31-atom supercell.}
\label{tab:NV_compilation}
\begin{ruledtabular}
\begin{tabular}{llccc}
Source & Method & $^1E$ & $^1A_1$ & $^3E$ \\
\hline
Experiment~\cite{daviesOpticalStudies19451976,rogersInfraredEmissionNV2008} & ZPL (PL/EPR) & $\sim 0.4$ (dark) & $\sim 1.6$ & 1.945 \\
This work & VMC, PBE0/CASCI/fixed Jast. & 0.91(1) & 2.60(2) & 3.04(1) \\
This work & VMC, fully optimized & 0.80(1) & 2.41(2) & 2.79(1) \\
This work & FN-DMC, fully optimized & 0.87(5) & 2.20(6) & 2.77(4) \\
\cite{maExcitedStatesNegatively2010} & GW+BSE, 256-atom cell & 0.40 & 0.99 & 2.32 \\
\cite{bockstedteInitioDescriptionHighly2018} & CI-CRPA, 512-atom cell & $\sim 0.5$ & $\sim 1.5$ & $\sim 2.0$ \\
\cite{bhandariMulticonfigurationalStudyNegatively2021} & CASSCF(6o,6e), 70-atom cluster & 0.34 & 1.41 & 1.93 \\
\cite{bhandariMulticonfigurationalStudyNegatively2021} & CASSCF(6o,6e), 162-atom cluster & 0.25 & 1.60 & 2.14 \\
\cite{muechlerQuantumEmbeddingMethods2022} & cRPA+ED, PBE+Hartree DC & 0.45 & 1.22 & 1.84 \\
\cite{shengGreensFunctionFormulation2022} & QDET, EDC@G$_0$W$_0$, 511-atom (22o,42e) & 0.463 & 1.270 & 2.152 \\
\cite{simulaCalculationEnergiesMultideterminant2023} & DMC, backflow, extrapolated & 1.0(1) & 2.2(1) & 2.43(9) \\
\end{tabular}
\end{ruledtabular}
\textit{Notes.} The $^3A_2\to{}^1E$ transition is optically dark; the
$^1A_1$ entry is the sum of the inferred dark gap ($\sim 0.4$~eV) and the
experimental $^1A_1\to{}^1E$ ZPL energy of $1.190$~eV~\cite{rogersInfraredEmissionNV2008}.
Bockstedte~et~al.\ values are estimated from the prose of
Ref.~\cite{bockstedteInitioDescriptionHighly2018}; precise values would need to be read from their Fig.~2.
Standard abbreviations: PL = photoluminescence; EPR = electron paramagnetic resonance; CASSCF(m, n) = complete active space self-consistent field theory with an active space of m orbitals and n electrons; GW = many-body perturbation theory in the GW approximation; BSE = Bethe–Salpeter equation; QDET = quantum defect embedding theory; CRPA = constrained random phase approximation; ED = exact diagonalization; DC = double counting; EDC = exact double counting.
\end{table*}

\begin{table*}
\caption{Vertical excitation energies (eV) for SiV$^0$:diamond, relative to
  the $^3A_{2g}$ ground state. This work's values are from a 31-atom supercell.}
\label{tab:SiV_compilation}
\begin{ruledtabular}
\begin{tabular}{llccc}
Source & Method & $^1E_g$ & $^1A_{1g}$ & $^3E_u$ \\
\hline
Experiment~\cite{greenElectronicStructureNeutral2019,dhaenens-johanssonOpticalPropertiesNeutral2011} & ZPL (PL/EPR) & --- (dark) & --- (dark) & 1.31 \\
This work & VMC, PBE0/CASCI/fixed Jast. & 0.42(1) & 0.87(2) & 3.21(1) \\
This work & VMC, fully optimized & 0.42(1) & 0.84(2) & 3.22(1) \\
This work & FN-DMC, fully optimized & 0.38(5) & 0.74(8) & 3.15(4) \\
\cite{galiInitioStudySplit2013} & HSE06 + constrained DFT & --- & --- & 1.63 \\
\cite{maFirstprinciplesStudiesStrongly2020} & QDET/FCI, PBE, beyond-RPA & 0.281 & 0.478 & 1.258 \\
\cite{maFirstprinciplesStudiesStrongly2020} & QDET/FCI, DDH, beyond-RPA & 0.336 & 0.583 & 1.594 \\
\cite{maQuantumEmbeddingTheory2021} & QDET/FCI, PBE, $W^E_{\mathrm{vel}}$ & 0.281 & 0.478 & 1.258 \\
\cite{maQuantumEmbeddingTheory2021} & QDET/FCI, DDH, $W^E_{\mathrm{vel}}$ & 0.336 & 0.583 & 1.594 \\
\end{tabular}
\end{ruledtabular}
\textit{Notes.} Singlet states of $^1E_g$ and $^1A_{1g}$ symmetry are not
accessible within constrained DFT and were not reported in
Ref.~\cite{galiInitioStudySplit2013}.
\end{table*}

\begin{table*}
\caption{Vertical excitation energies (eV) for Fe$_{\mathrm{Al}}^0$:AlN, relative
  to the $^6A_1$ ground state. This work's values are from a 32-atom supercell.
  In tetrahedral notation, the lower $^4A_2+{}^4E$ pair derives from $^4T_1$
  and the upper $^4A_1+{}^4E$ pair from $^4T_2$.}
\label{tab:FeAlN_compilation}
\begin{ruledtabular}
\begin{tabular}{llcccc}
Source & Method & $^4A_2$ & $^4E$ (lower) & $^4A_1$ & $^4E$ (upper) \\
\hline
Experiment~\cite{baurDeterminationGaNAlN1994,masendaLatticeSitesCharge2016} & PL / M\"ossbauer & $\sim 1.30$ & $\sim 1.30$ & $\sim 1.78$ & $\sim 1.78$ \\
This work & VMC, PBE0/CASCI/fixed Jast. & 1.70(5) & 1.74(4) & 2.30(5) & 2.36(3) \\
This work & VMC, fully optimized & 1.59(3) & 1.68(2) & 2.20(3) & 2.29(2) \\
This work & FN-DMC, fully optimized & 1.5(1) & 1.69(9) & 2.3(1) & 2.27(9) \\
\cite{kleinerQuantumMonteCarlo2025} & VMC, HSE06/vqz orbitals & 1.70(6) & 1.78(4) & 2.27(5) & 2.35(4) \\
\cite{otisStronglyCorrelatedStates2025} & CAS-DMET, (15e,15o) & --- & 1.94 & --- & --- \\
\cite{otisStronglyCorrelatedStates2025} & NEVPT2-DMET, (15e,15o) & --- & 1.46 & --- & --- \\
\cite{otisStronglyCorrelatedStates2025} & PBE-QDET, (5e,5o) & --- & 1.02 & --- & --- \\
\cite{otisStronglyCorrelatedStates2025} & HSE-QDET, (5e,5o) & --- & 1.00 & --- & --- \\
\cite{otisStronglyCorrelatedStates2025} & HSE-TDDFT & --- & 1.50 & --- & --- \\
\end{tabular}
\end{ruledtabular}
\textit{Notes.} Experimental PL bands are reported as broad features that
do not resolve the $^4A_2$/$^4E$ or $^4A_1$/$^4E$ splittings; the same
band-center value is therefore listed in both columns of each pair. The
Otis~et~al.\ entries are adiabatic excitation energies (ZPL); only
the lowest $^4E$ state is reported in
Ref.~\cite{otisStronglyCorrelatedStates2025}. The cRPA+ED embedding of
Ref.~\cite{muechlerQuantumEmbeddingMethods2022} yields an incorrect state
ordering for this defect and is omitted here.
Standard abbreviations: DMET = density matrix embedding theory; NEVPT2 = $N$-electron valence state second-order perturbation theory; TDDFT = time-dependent DFT.
\end{table*}

\begin{table*}
\caption{Vertical excitation energies (eV) for Cr$_{\mathrm{Al}}^+$:AlN, relative
  to the $^3A_2$ ground state. This work's values are from a 32-atom supercell.
  Czelej et~al.~\cite{czelejTransitionMetalRelatedQuantumEmitters2024}\ report the symmetry assignments using tetrahedral notation; their $^3T_2$ would
  split into $^3E + {}^3A_1$ in $C_{3v}$.}
\label{tab:CrAlN_compilation}
\begin{ruledtabular}
\begin{tabular}{llcccc}
Source & Method & $^1E$ & $^3E$ & $^3A_1$ & $^1A_1$ \\
\hline
Experiment~\cite{baurPhotoluminescenceResidualTransition1995} & ZPL (PL) & --- (spin-forbidden) & $\sim 1.21$ & --- & --- (spin-forbidden) \\
This work & VMC, PBE0/CASCI/fixed Jast. & 1.41(2) & 2.50(2) & 2.67(3) & 2.71(2) \\
This work & VMC, fully optimized & 1.34(1) & 2.13(1) & 2.33(2) & 2.59(2) \\
This work & FN-DMC, fully optimized & 1.50(5) & 2.03(6) & 2.16(8) & 2.78(8) \\
\cite{czelejTransitionMetalRelatedQuantumEmitters2024} & mcDFT (HSE constrained-occupation) & 1.25 & 1.83 & 1.83 & --- \\
\cite{czelejTransitionMetalRelatedQuantumEmitters2024} & CASCI/NEVPT2 (cluster) & 1.26 & 1.92 & 1.92 & --- \\
\end{tabular}
\end{ruledtabular}
\textit{Notes.} The experimental PL lines near 1.2~eV are attributed
specifically to Cr$_{\mathrm{Al}}^+$ in
Ref.~\cite{czelejTransitionMetalRelatedQuantumEmitters2024}, which reports
states in $T_d$ notation; the listed value is the $^3T_2$ energy and is
shown in both the $^3E$ and $^3A_1$ columns since these two irreps emerge
together from $^3T_2$ in $C_{3v}$ symmetry.
Standard abbreviations: mcDFT = multi-configurational DFT.
\end{table*}

\bibliography{references}

\end{document}